\documentclass{cjaa-1} 
\usepackage{graphicx}                   
\input{epsf.sty}                        
\input{psfig.sty}                       

\begin{document}

   \title{Timescale Analysis of Spectral Lags
}

   \volnopage{Vol.0 (200x) No.0, 000--000}      
   \setcounter{page}{1}          

\author{Ti-Pei Li
      \inst{1,2,3}\mailto{}
   \and Jin-Lu Qu
      \inst{2}
   \and Hua Feng
      \inst{3}
   \and Li-Ming Song
      \inst{2}
   \and \\ Guo-Qiang Ding
      \inst{2} 
   \and Li Chen
      \inst{4}
      }
   \offprints{T.P. Li}                   

\institute{Department of Physics \& Center for Astrophysics, Tsinghua
University, Beijing\\
 \email{litp@mail.tsinghua.edu.cn}    
   \and
Particle Astrophysics Lab., Inst, of High Energy Physics, Chinese Academy of
Sciences       
   \and
Department of Engineering Physics \& Center for Astrophysics, Tsinghua
University
   \and
Department of Astronomy, Beijing Normal University       
   }

   \date{Accepted 2004 June 18}

   \abstract{A technique for timescale analysis of spectral lags performed directly 
in the time domain is developed. Simulation studies are made
to compare the time domain technique with the Fourier frequency analysis
for spectral time lags.  The time domain technique is applied  to studying 
rapid variabilities of X-ray binaries and $\gamma$-ray bursts.
 The results indicate that in comparison with the Fourier analysis the timescale 
analysis technique is more powerful for the study of spectral lags in rapid 
variabilities on short time scales and short duration flaring phenomena.
\keywords{methods: data analysis --- binaries: general --- X-rays: stars --- gamma rays: bursts --- X-rays: bursts}
   }

   \authorrunning{T. P. Li, H. Feng, L. M. Song, G. Q. Ding \& L. Chen }            
   \titlerunning{Timescale Analysis of Spectral Lags }  

   \maketitle

%
%
\section{Introduction} 
The analysis of spectral lag between variation signals in different energy bands
is an important approach to get useful information on  their producing and 
propagation processes in celestial objects. Observed intensity variations 
are usually produced by various processes with different time scales and different 
spectral lags. 
A lag spectrum, a distribution of time lags over Fourier frequencies, 
can be derived from two related time series with the aid of the Fourier 
transformation.    
Let $x(t_i)$ and $y(t_i)$ be two light curves observed simultaneously in 
two energy bands at times $t_i$, their Fourier transforms are 
$X(f_j)$ and $Y(f_j)$ respectively, and the cross spectrum 
$C(f_j)=X^*(f_j)Y(f_j)$. The argument of the cross 
spectrum $C(f_j)$ is the phase difference between  the two processes at 
frequency $f_j$, or the time lag of photons in band 2 relative to that 
in band~1 
\begin{equation} \tau (f_j)=\arg [C(f_j)]/2\pi f_j ~.\label{cf} \end{equation} 
The Fourier analysis technique has been most widely used in studying 
spectral lags.

The time domain method for studying spectral lags can be based on 
the correlation analysis.  For two counting series $x(t_i),~ y(t_i)$ (or 
$x(i),~ y(i)$ ), the observed counts in the corresponding energy band 
in the time interval $(t_i, t_{i+1})$ with $t_i=(i-1)\Delta t$,    
 the cross-correlation function (CCF) of the zero-mean time series 
at lag $k\Delta t$ is usually defined as
\begin{equation} \mbox{CCF}(k)=\sum_i u(i)v(i+k)/\sigma(u)\sigma(v) 
\hspace{3mm} (k=1,\pm 1,\cdots) \label{ccf}\end{equation}
with $u(i)=x(i)-\bar{x}, v(i)=y(i)-\bar{y}$, $\sigma^2(u)=\sum_i[u(i)]^2$ 
and  $\sigma^2(v)=\sum_i[v(i)]^2$. With CCF the time lag 
can be defined as $\Lambda = k_m\Delta t$ where CCF($k$)/CCF(0) has maximum 
at $k=k_m$. 
Instead of a lag spectrum provided by Fourier
analysis, the correlation technique gives only a single value
$\Lambda$ of time lag.
To an understanding of a physical process occurring 
in the time domain, we need to know spectral lags 
at different timescales, i.e. a timescale spectrum $\Lambda(\Delta t)$.   

We can not simply equate a Fourier period with the timescale and interpret 
a Fourier spectrum in the time domain as the timescale spectrum. 
For example, a Fourier power spectrum can not be interpreted 
as the distribution of variability amplitude vs. timescale. 
A sinusoidal process with frequency $f$ 
has no Fourier power at any frequency except $f$, but it does not mean 
that no variation exists at timescales shorter than $1/f$.
One can make lightcurves with time steps smaller than $1/f$ and find that 
non-Poissonian variations of intensities do exist in such lightcurves.
In fact, a frequency analysis is based on a certain kind of time-frequency 
transformation. 
Different mathematically equivalent representations with different bases or 
functional coordinates in the frequency domain exist for a certain time series,
a Fourier spectrum with the trigonometric basis does not necessarily 
represent the true distribution of a physical process in the time domain.   
It has to be kept in mind that a mathematical transform
may distort physical information contained in the observational data. 
For correct understanding the real process, one has to invert 
results obtained
through a time-frequency transform into the real physical space.
It is usually not easy to complete such an inversion. 
A sinusoidal process is the simplest signal in the 
frequency domain, but a complex one in the time domain. 
The correct procedure to invert 
a Fourier power spectrum $p_{_F}(f)$ into the timescale
spectrum $p(\Delta t)$ in the physical space (the time domain) is
\[ p(\Delta t)=\int p_{_F}(f)~ p(\Delta t|f)~ \Delta t~, 
\]
where $p(\Delta t|f)$ is the timescale spectrum of a sinusoidal process
with frequency $f$ and unit amplitude, which is not a simple value or function 
and can not be derived from the Fourier analysis.        

To understand a time process correctly, we have to make timescale analysis directly
in the time domain and need to develop spectral analysis technique in the time domain
without using the Fourier transform or other time-frequency transformation. 
A preliminary algorithm to modify the conventional cross-correlation technique 
was proposed by Li, Feng \& Chen (1999). 
After then the algorithms to evaluate timescale spectra
of power density, coherence, spectral hardness, variability duration, 
and correlation coefficient between two characteristic quantities 
were worked out (\cite{lit01}), the modified 
cross-correlation technique is a part of the timescale analysis method 
in the time domain. Recently we have developed and completed the modified cross-correlation 
technique, improved its sensitivity and lag resolution significantly. 
This paper presents the timescale analysis technique of spectral 
lags and its application to analyzing space hard x-ray and $\gamma$-ray data. 
The general procedure of timescale analyzing and the modified cross-correlation 
function for spectral lag analysis in the time domain  are introduced 
in \S 2. The technique has been applied to studying spectral lags of hard X-rays
from X-ray binaries, $\gamma$-ray bursts and terrestrial $\gamma$-ray flashes,
some examples are shown in \S 3. Relevant discussions are made in \S 4.  

\section{Method}  
\subsection{Timescale Analysis}
Temporal analysis is an important approach to study dynamics of 
physical processes in objects.  Usually we take some quantities, e.g., 
power density (variation amplitude), spectral lag, and coherence etc.,
to characterize temporal property of observed light curves.  
The complex variability of high-energy emission shown in different
time scales is a common character for X-ray binaries, super massive black
 holes and $\gamma$-ray bursts. The variability is caused by various 
physical processes at different timescales. 
 It is not easy to study 
the variation phenomena on a given time scale. Large time bin used 
in calculation will erase the information on shorter time scales. 
And the analysis result with a short time bin reflects not only 
the variation property on the short time scale, but may also be affected 
by that on longer ones up to the total time period used in the calculation.  
The fact that a lightcurve with time step $\Delta t$ does not include 
any information of variabilities at timescales shorter than $\Delta t$ 
can be used as
a foundation of timescale analysis. A set of lightcurves with different
time steps $\Delta t$ produced by rebinning the same originally observed
data with a time resolution $\delta t$ is the basic material 
in timescale analysis.

Usually the originally observed data for temporal analysis is a counting 
series  $x(j; \delta t) ~(j=1,\cdots)$ with a time resolution $\delta t$.
 To study variability on a timescale 
\begin{equation}
\Delta t = M_{\Delta t}\delta t ~,\end{equation}
we need to construct a new lightcurve with the time step $\Delta t$ from the native series by
combining its $M_{\Delta t}$ successive bins by
\begin{equation} 
x(i; \Delta t)=\sum_{j=(i-1)M_{\Delta t}+1}^{iM_{\Delta t}}x(j; \delta t) ~. 
\end{equation}
As the lightcurve $\{x(\Delta t)\}$ does not include any information
about the variation on any timescale shorter than $\Delta t$, it is suitable
for studying variability over the region of timescale $\ge \Delta t$. 
 
Let $\Lambda$ denote the quantity under study. The value $\Lambda(\Delta t)$ 
of the quantity at the timescale $\Delta t$ can be seen as a function
of the lightcurve $x(i; \Delta t)$
\begin{equation} \Lambda(\Delta t)=f_{\Lambda}[\{x(\Delta t)\}]~. \end{equation}
The key point in timescale analysis is to find a proper algorithm $f_{\Lambda}$ to calculate
the value of the studied quantity $\Lambda$ at a certain timescale $\Delta t$. 

\begin{figure}
\hspace{-1cm}
 \includegraphics[width=120mm]{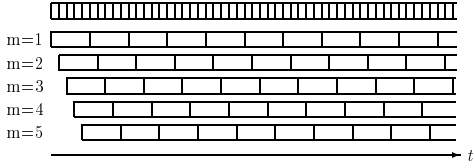}
\vspace{-13cm}
 \caption{From an originally observed time series with time resolution $\delta t$
(schematically shown at the top where each small square represents a time bin with width
$\delta t$) five different lightcurves with time step $\Delta t=5\delta t$ can be constructed
with different phase parameter $m$.
\label{fig:lcs}}
  \end{figure}

The procedure (4) of binning the native series $\{x(\delta t)\}$ to get $\{x(\Delta t)\}$
with a larger time step $ \Delta t = M_{\Delta t}\delta t$ is started from the first
bin of $x(j=1; \delta t)$. From the native lightcurve we can get $M_{\Delta t}$ different
lightcurves with the same time step $\Delta t$ 
\begin{equation} 
x_{m}(i; \Delta t)=\sum_{j=(i-1)M_{\Delta t}+m}^{iM_{\Delta t}+m-1}x(j; \delta t) ~,
\end{equation}
where the combination starts from the $m$th bin of the native
series, the phase factor $m=1,\cdots,M_{\Delta t}$ (see the diagrammatic sketch
Fig. \ref{fig:lcs}).

For sufficiently using the information about variation on the timescale $\Delta t$ 
included in the originally observed lightcurve, we can calculate the studied quantity 
$\Lambda(\Delta t)$ for each $\{x_{m}(\Delta t)\}$ and take their average as the resultant
value
\begin{equation}
 \Lambda(\Delta t)=\frac{1}{M_{\Delta t}}\sum_{m=1}^{M_{\Delta t}}f_{\Lambda}[\{x_{m}(\Delta t)\}]~. 
\end{equation}
 
In timescale analysis the observed lightcurve $\{x(\delta t)\}$ is usually divided into $L$ segments, 
each segment includes nearly equal number of successive bins. 
For each segment~$i$, we can get a value $\Lambda_i(\Delta t)$ by Eq. (7),
the average $\overline{\Lambda}(\Delta t)$ and
its standard deviation $\sigma (\overline{\Lambda}(\Delta t)$ can be derived
\begin{eqnarray}  \overline{\Lambda}(\Delta t)&=&\sum_{i=1}^L
\Lambda_i(\Delta t)/L ~, \nonumber \\ 
\sigma (\overline{\Lambda}(\Delta t))&=&\sqrt{\sum_{i=1}^{L}(\Lambda_i(\Delta t)-
\overline{\Lambda}(\Delta t))^2/L(L-1)} ~.
 \end{eqnarray}

 Usually we can use some convenient statistical methods based on
the normal distribution to make statistical inference, e.g. significance test,
on $\overline{\Lambda}(\Delta t)$. For the case of short time scale  $\Delta t$, although the number  
of counts per bin may be too small for it to be assumed as a normal variable, it is easy
from a certain observation period to get the total number $L$ of segments large enough 
to satisfy the condition for applying the central limit theorem in statistics    
and using the normal statistics for the mean $\overline{\Lambda}(\Delta t)$. 

\subsection{Modified Cross Correlation Function}
In timescale analysis for spectral lags, the observed data are two related counting series
$\{x(i; \delta t)\}$ and $\{y(i; \delta t)\}$ in two energy bands with time resolution 
$\delta t$. If the timescale $\Delta t$ under study is larger than the time resolution,
we need to construct two new lightcurves, $\{x(i; \Delta t)\}$ and $\{y(i; \Delta t)\}$,
by re-binning the originally observed series with the time step $\Delta t$.
With the traditional CCF defined by Eq. (\ref{ccf}), we can calculate the time lag $\Lambda$
only if $\Lambda > \Delta t$. But for many physical processes the real
time lag is shorter or even much shorter than the process timescale. For the purpose of applying 
correlation analysis to
the general case of lag analysis, a modified CCF at lag $k\delta t$ has been proposed (\cite{lit99}) 
\begin{equation}
\mbox{MCCF}_0(k; \Delta t)=\sum_iu_1(i; \Delta t) v_{k+1}(i; \Delta t)
/\sigma(u)\sigma(v) ~,
\end{equation}
where the time step $\Delta t=M_{\Delta t}\delta t$, $\{u_k(\Delta t)\}$ and $\{v_k(\Delta t)\}$ 
are the zero-mean series of $\{x_k(\Delta t)\}$ and
$\{y_k(\Delta t)\}$ respectively, and
\begin{eqnarray} 
x_{m}(i; \Delta t)&=&\sum_{j=(i-1)M_{\Delta t}+m}^{iM_{\Delta t}+m-1}x(j; \delta t) ~, \nonumber\\
y_{m}(i; \Delta t)&=&\sum_{j=(i-1)M_{\Delta t}+m}^{iM_{\Delta t}+m-1}y(j; \delta t) ~.
\end{eqnarray}
The lag resolution of CCF defined by Eq. (2) is $\Delta t$, and that of MCCF$_0$ defined above 
is the original time resolution $\delta t$. 
 
\begin{figure}
\hspace{-1.7cm}
 \includegraphics[width=120mm]{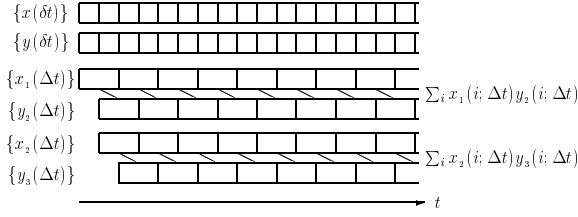}
\vspace{-13cm}
 \caption{MCCF of two native time series $\{x(\delta t)\}$ 
and $\{y(\delta t)\}$
at a time lag $\tau=\delta t~(\tau=k\delta t,~ k=1)$ 
and on a timescale $\Delta t = 2 \delta t$. 
\( \mbox{MCCF}(k=1; \Delta t)=\frac{1}{2}[\sum
_ix_{_1}(i; \Delta t) y_{_2}(i; \Delta t) + \sum_ix_{_2}(i; \Delta t) 
y_{_3}(i; \Delta t)] \)
\label{fig:cc}}
\end{figure}

For sufficiently using the information contained in the observed lightcurve, we propose 
to improve the definition of MCCF$_0$ further by following the procedure described by Eq. (7).
The new and complete definition of MCCF at lag $k\delta t$ is
\begin{eqnarray}
\lefteqn{\mbox{MCCF}(k; \Delta t)=} \nonumber \\
 & & \frac{1}{M_{\Delta t}}\sum_{m=1}^{M_{\Delta t}}
\sum_{i}u_{m}(i; \Delta t) v_{m+k}(i; \Delta t)/\sigma(u)\sigma(v) \label{mccf}
\end{eqnarray} 
where $u_m(i; \Delta t)=x_m(i; \Delta t)-\bar{x}_m(\Delta t),~v_m(i; \Delta t)=
y_m(i; \Delta t)-\bar{y}_m(\Delta t)$, $\bar{x}_m(\Delta t)$ and 
$\bar{y}_m(\Delta t)$ are the current averages for the used segments 
of lightcurve $\{x\}$  and $\{y\}$ respectively.
The procedure of calculating a modified cross-correlation coefficient 
is schematically shown by Fig. \ref{fig:cc}.  
We can find a value $k_m$ of $k$ to satisfy the condition
\begin{equation} \mbox{MCCF}(k=k_m; \Delta t)/\mbox{MCCF}(0; \Delta t)=\max~, 
\end{equation} 
then the lag of band 2 relative to band 1 on timescale
$\Delta t$
\begin{equation}
\Lambda(\Delta t)=k_m\delta t~.
\end{equation}

For an observed time series $x(j; \delta t)$ with time resolution $\delta t$, we can similarly
define a modified auto-correlation function at lag $k\delta t$ on timescale 
$\Delta t=M_{\Delta t}\delta t$
\begin{eqnarray}
\lefteqn{\mbox{MACF}(k; \Delta t)=} \nonumber \\
 & &\frac{1}{M_{\Delta t}}\sum_{m=1}^{M_{\Delta t}}
\sum_{i}u_{m}(i; \Delta t) u_{m+k}(i; \Delta t)/\sigma^2(u)
\end{eqnarray}
where the $u_{m}(i; \Delta t)=x_m(i; \Delta t)-\bar{x}$, $x_{m}(i; \Delta t)=
\sum_{j=(i-1)M_{\Delta t}+m}^{iM_{\Delta t}+m-1}x(j; \delta t)$.
The FWHM of MACF$(\tau; \Delta t)$ can be taken as a measure of the duration of variation 
on time scale $\Delta t$. With MACF
we can study the energy dependence of average shot width in a random shot process
on different time scales.  
     
\subsection{Simulations}
\subsubsection{Poissonian signals}
\begin{figure}
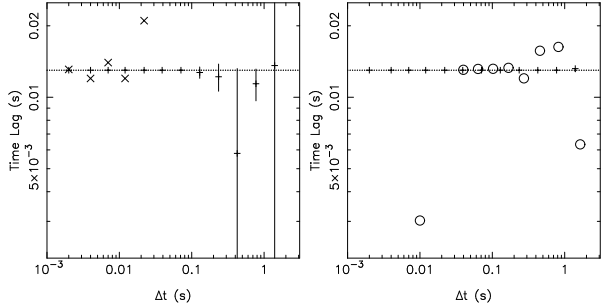

 \includegraphics[width=40mm,angle=270]{f3a.eps}
 \includegraphics[width=40mm,angle=270]{f3b.eps}
 \caption{Time lag vs. time scale of two white noise series with 13 ms time lag
shown by the dotted horizontal line. 
{\sl Left panel}: {\it Cross} -- CCF lag; {\it Plus} -- lag evaluated by 
MCCF$_0$ [Eq.(9)].
{\sl Right panel}: {\it Circle} -- lag from Fourier analysis; {\it Plus} -- 
lag by MCCF [Eq.(11)].
\label{fig:testcc}}
\end{figure}

To compare the above MCCF technique of estimating time lags with the
traditional CCF technique and Fourier analysis, we produce two 
photon event series of length 1000~s with a known time lag between them.
The series~1 is a white noise series with average rate 200 cts s$^{-1}$ 
and series~2 consists of the same events in series 1 but each event time 
is delayed 13~ms. 
Besides the signal photons mentioned above, the two series are given 
independent additional noise events at average rate 300 cts s$^{-1}$. 
By binning the two event series, two light curves with time resolution 
$\delta t=1$ ms are produced. 
We make time lag analysis at  timescales $\Delta t$ from 1 ms to 2 s
for the two lightcurves by  CCF, MCCF and Fourier analysis techniques separately 
(in Fourier analysis  we use Fourier cross spectrum with 1~ms light curves 
and 4096-point FFT and take Fourier frequency $f=1/\Delta t$), the results 
are shown in Fig. \ref{fig:testcc}. 
In the left panel of Fig. \ref{fig:testcc},  the cross signs show lags derived by CCF and
plus signs by MCCF$_0$ defined by Eq. (9).  For the timescale region of $\Delta t$ shorter or approximately
equal to the magnitude of the true lag 0.013 s where CCF works,
MCCF$_0$ can provide more reliable results with better accuracy.
The circles in the right panel of Fig. \ref{fig:testcc} indicate the Fourier lags, 
for the short timescale region of $\Delta t \la 0.3$~s (or high frequency region
of $f \ga 30$ Hz) the Fourier analysis can not give any meaningful result.

In the timescale region of $\Delta t \ga 0.1$ s, the estimates of time lag by MCCF$_0$
defined by Eq.~(9) 
(plus signs in the left panel of Fig. \ref{fig:testcc}) show significant fluctuation about 
the expectation. 
We use the improved MCCF defined by Eq.~(11) to calculate the lag spectrum again, 
the plus signs in the right 
panel of Fig. \ref{fig:testcc} indicate the result. Comparing the lag spectra with MCCF$_0$ [Eq. (9)] and
MCCF [Eq.~(11)], plus signs in the left panel and right panel of Fig. \ref{fig:testcc},
we can see that using the improved MCCF can improve the lag spectrum significantly
in the large timescale region.    

\subsubsection{Transient signals}
\begin{figure}
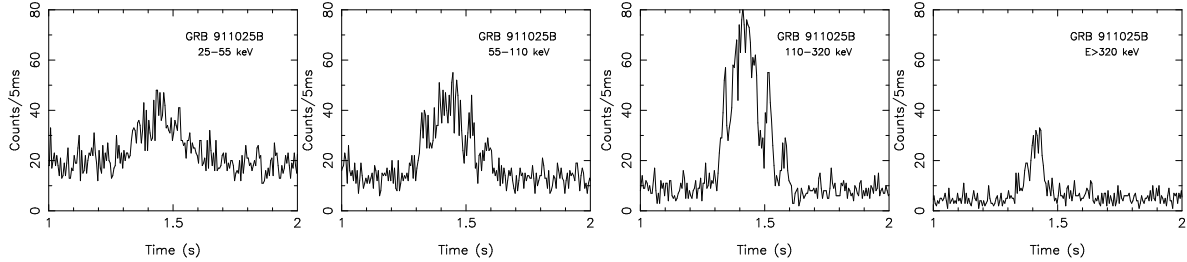

\includegraphics[width=34mm,angle=270]{f4a.eps}\includegraphics[width=34mm,angle=270]{f4b.eps}
\includegraphics[width=34mm,angle=270]{f4c.eps}\includegraphics[width=34mm,angle=270]{f4d.eps}
 \caption{Light curves of a $\gamma$-ray burst GRB 911025B observed 
by BATSE in channels 25-55 keV, 55-110 keV, 110-320 keV, and $>320$ keV,
separately.
\label{fig:grblc}}
\end{figure}

\begin{figure}
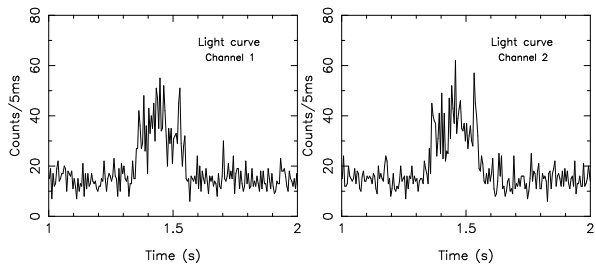

\includegraphics[width=34mm,angle=270]{f5a.eps}\includegraphics[width=34mm,angle=270]{f5b.eps}
 \caption{Simulated light curves for two channels. The burst process for 
channel 2 is delayed 0.01 s to channel 1.  
\label{fig:mclc}}
\end{figure}

 To study the relative timing of transient emission 
at different energies is a difficult task in astrophysics. 
Neither the Fourier analysis nor the traditional cross-correlation technique
can get meaningful result of spectral lags from the observed data of
short $\gamma$-ray bursts.
We show here the ability of MCCF to study spectral lags
in prompt emission of short $\gamma$-ray bursts by simulation.
Fig. \ref{fig:grblc} shows light curves with 5 ms time bin observed 
by BATSE on $CGRO$ mission
for a $\gamma$-ray burst, GRB 911025B  (BATSE trigger number 936), in channel
25-55 keV, 55-110 keV, 110-320 keV, and $>320$ keV, respectively. 
The time-tagged event (TTE) data of BATSE
contains the arrival time ($2\mu s$ resolution) of each photon
for the short duration burst GRB 911025B. 
In our simulation, we use the 25-55 keV photons  
between 1.35~s and 1.55~s in the TTE data as the burst events
in channel~1. The burst events in channel~2 are the same in channel~1
but each time is delayed 0.01~s. By binning the two event series 
with time bin 5~ms, expected signal series in channel~1 and 2 
are produced.
Two simulated lightcurves in channel~1 and 2 are generated by taking
random samples from the expected signal series and adding independent 
background noise at the average of 15 counts each bin, 
shown in Fig. \ref{fig:mclc}.       

\begin{figure}
\hspace{1.5cm}
 \includegraphics[width=38mm,angle=270]{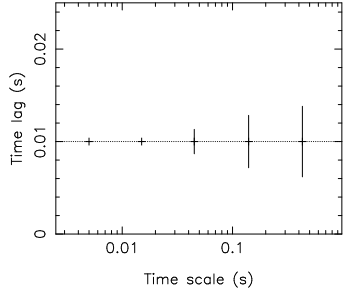}
 \caption{Time lag vs timescale of two simulated light curves 
in Fig. \ref{fig:mclc}.
{\it Dotted line} -- expected time lag.
{\it Cross} -- measured by MCCF.  
\label{fig:grblag}}
\end{figure}

From the two simulated lightcurves 
we use MCCF to calculate the time lags at timescale 
$\Delta t=0.005$~s, 0.015~s, 0.045~s, 0.14~s, and 0.43~s separately, the results are shown in 
Fig. \ref{fig:grblag}. The error bar of time lag at each timescale
is estimated from 200 bootstrap samples.  The simulation result indicates
that MCCF is a useful tool in relative timing of transient processes.
 
\subsubsection{Timescale dependent process}
\begin{figure}
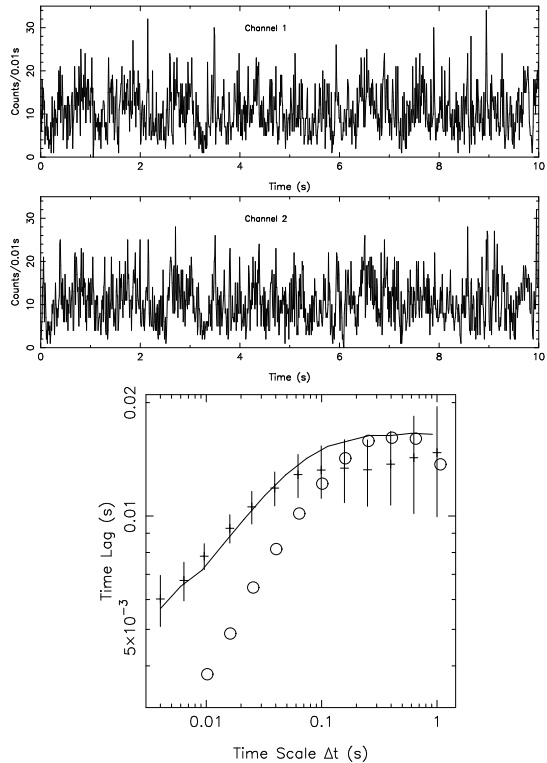

\begin{center}
\includegraphics[width=25mm,angle=270]{f7a.eps}\\  
\includegraphics[width=25mm,angle=270]{f7b.eps}\\ 
\includegraphics[width=50mm,angle=270]{f7c.eps}
\end{center} 
 \caption{{\sl Top and Middle}: Synthetic lightcurves with time bin 0.01 s 
for a complex process 
consisting of different characteristic timescales and different spectral 
lags.
{\sl Bottom}: Timescale spectra of time lag; 
{\it Solid line} -- the expected lag spectrum,
{\it Plus} -- MCCF lag,
{\it Circle} -- lag from Fourier analysis. 
\label{fig:com}}
\end{figure}

Spectral time lags observed for X-ray binaries and AGNs are usually 
timescale dependent. For example, the Fourier lag between 14.1-45 keV 
and 0-3.9 keV of X-ray emission from Cyg X-1 in the low state continuously
varies with Fourier frequency, from $\sim 30$~ms at 0.1~Hz decreasing down to
$\sim 2$~ms at 10~Hz (\cite{now99}). Real data in timing can be seen as
a complex time series with multiple timescale components.
The timescale dependence of spectral lags
can be the intrinsic property of the emission process and/or 
come from different processes dominating at different timescales.
Correctly detecting the timescale dependence of spectral lags
is important to studying the undergoing physical processes. 
Now we compare the abilities of MCCF and Fourier technique to study 
spectral lags of time series with multiple timescale components.
Two light curves of a complex process consisting of five independent 
random shot components are produced by Monte Carlo simulation. 
Each signal component $i~(i=1-5)$ consists of random 
shots with shape of profile $a\cdot\exp [(t-t_0)/\tau_i)]^2$, where the peak height $a$ is
randomly taken from the uniform distribution between zero and the maximum. 
The separation between two successive shots is exponentially distributed with  
average separation $\tau_i$. For each component $i$, we produce a 3000 s counting series 
of band 1 with a step of $\delta t=1$ ms and average rate 200 cts s$^{-1}$. 
The corresponding series in band 2 consists of the same events in series 1 but
each event time is delayed $\Lambda_i$ s. The characteristic time constants 
$\tau_i$ of the five signal components are 0.005, 0.01, 0.02, 0.04, and 0.08 s, and their time lags are 0.004, 0.008, 0.012, 0.016, and 0.02 s
respectively.  Summing up the five series for each band, we produce two expected 
signal lightcurves. Two synthetic light curves with time step 1 ms are made by
random sampling the expected light curves with Poisson fluctuation plus a
independent white noise at mean rate of 100 cts/ s$^{-1}$. For the two light curves,
we use their Fourier cross spectrum and 4096-point FFT and MCCF in the time domain
to calculate the time lags on different time scales $\Delta t$ and show the
results in Fig. \ref{fig:com}. 

\begin{figure}
\hspace{1cm}
 \includegraphics[width=40mm,angle=270]{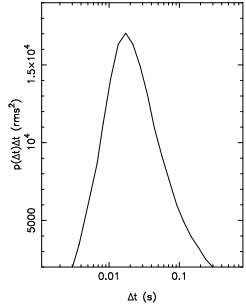} 
 \caption{Power distribution $p(\Delta t)\Delta t$ vs. time scale $\Delta t$ of
a random shot process with a characteristic timescale $\tau=0.01$ s, where
$p(\Delta t)$ (rms$^2$/s) is the power density on the timescale $\Delta t$
of the process. 
\label{fig:pow}}
\end{figure}

To derive the expected lag spectrum of the above synthetic lightcurves,
we calculate the timescale distribution of power density for each shot component
using the algorithm of estimating power density spectrum $p(\Delta t)$ 
in the time domain (\cite{lit01}; \cite{lit02}). 
As an example, Fig. \ref{fig:pow} shows 
the distribution of variation power $p(\Delta t)\Delta t$ (rms$^2$) 
vs. time scale $\Delta t$ for the expected light curve of the shot process with 
$\tau=0.01$ s.  As shown in Fig. \ref{fig:pow}, an individual shot component with 
characteristic time $\tau_i$ has 
its variation power distributed over a certain time scale region. 
The time lag between two bands should appear  
in the whole timescale region where the signal variation power exists. 
The expected lag of the two synthetic light curves with multiple components 
on timescale $\Delta t$
should be estimated as a weighted average of five lags $\Lambda_i$
of individual component with corresponding weight factor $p_i(\Delta t)\Delta t $   
\begin{equation}
\Lambda(\Delta t)=\sum_ip_i(\Delta t)\Delta t\cdot\Lambda_i \end{equation}
The solid line in Fig. \ref{fig:com} is the expected lag distribution calculated by Eq. (15).
We can see from Fig. \ref{fig:com} that the Fourier cross spectrum fails to detect lags
in the short timescale region, but MCCF works well. More simulations with different
signal to noise ratios show that the MCCF technique is capable of correctly detecting
time lags from severely noisy data and the inefficiency of detecting
time lag in short time scale region is an intrinsic weakness of the Fourier technique,
 even for data with much higher signal to noise ratio  the Fourier analysis still
can not detect lags in the high frequency region.

\section{Applications} 
The timescale spectral method for time lag analysis is a powerful tool 
in revealing the characteristic of emission process in objects.
With the help of MCCF technique, as a example, we can judge between
different production models of x-rays from accreting black holes.   
The energy spectra of hard X-rays from black hole binaries
can be fitted well by Comptonization of soft photons by hot electrons 
in the vicinity of the compact sources. To explain the observed energy 
spectra  the uniform corona model was suggested initially (\cite{pay80}), 
in which the soft photons from the central region 
of the system are Comptonized by the hot electrons of corona. 
The Comptonization process makes the observed hard photons undergo more 
scattering than the low energy photons and therefore the hard photons
are naturally delayed from soft photons. Hard lags are not correlated with
the variability timescale (or variability frequency) in the uniform
corona model, but later study show strong timescale-dependence of time lags 
(\cite{miy88}). For overcoming the contradiction 
between prediction by the uniform corona model and
observed results, other models, such as the non-uniform corona
model (\cite{kaz97}), the magnetic flare model (\cite{pou99}) 
and the drifting-blob model (\cite{bot99}) are proposed.
The hard X-ray lags are studied by observations with PCA detector 
on board $RXTE$ mission in using the Fourier technique to the black hole
candidate Cyg X-1 in the low 
state (\cite{now99}), high state (\cite{cui97a}), and during spectra 
transitions (\cite{cui97b}).
The meaningful Fourier spectra of time lag from PCA/$RXTE$ data are 
all limited in the range of Fourier frequency $\la 30$ Hz 
(or timescale $\ga 0.03$ s) and, except that with a uniform corona, 
all models mentioned above can fit the observed lag spectra of Cyg X-1. 
To test these models, we need to compare the 
expected and observed lags in the higher frequency range or on the shorter time scales.  
    
\begin{figure}
\includegraphics[width=32mm,angle=270]{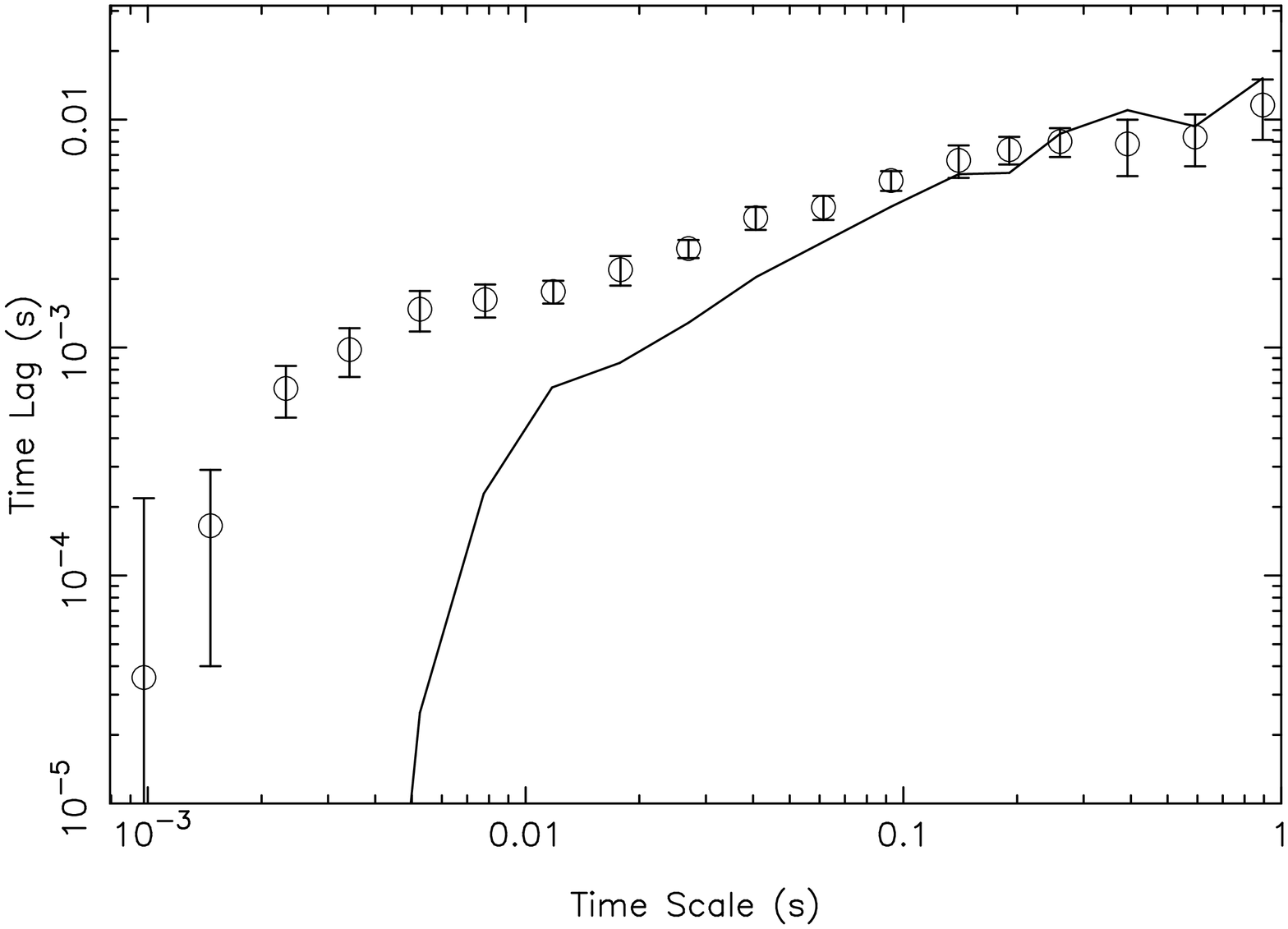}
\includegraphics[width=32mm,angle=270]{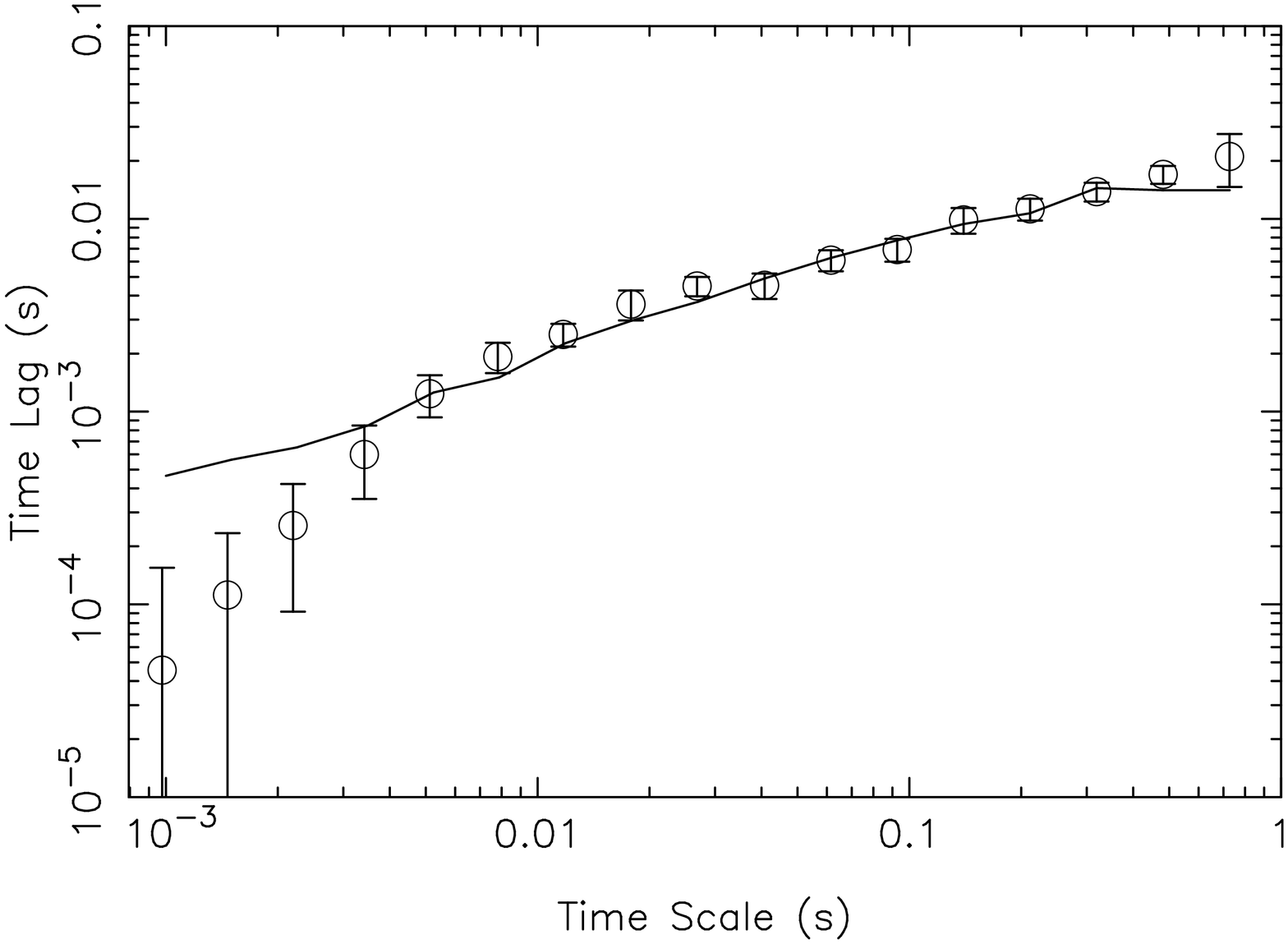}
 \caption{Hard X-ray time lag vs. time scale of Cyg X-1.
{\sl Left panel}: {\it Circle} -- time lag between 13 - 60 keV and 2 - 5 keV in the soft state of
Cyg X-1 on 1996 June 17 (PCA/$RXTE$ ObsID P10512) measured by MCCF; {\it Solid line} -- expected by the 
non-uniform corona model. 
{\sl Right panel}: {\it Circle} -- lags between 13 - 60 keV and 2 - 5 keV of
Cyg X-1 in the hard state on 1996 October 23 (PCA/$RXTE$ ObsID P10241) measured by MCCF; 
{\it Solid line} -- lags between 27 keV and 3 keV predicted  
by the magnetic flare model.
\label{fig:cyg}}
\end{figure}

Kazanas et al (1997) present that for the PCA/$RXTE$ observation of 
June 16 1996 (ObsID P10512), hard X-ray time lags of Cyg X-1 in the soft state 
as a function of Fourier frequency 
over the region of 8 - $\sim$30 Hz can be well fitted by the non-uniform corona model.
From the same data, we measure the time lags between 13-60 keV and 2-5 keV 
 on short time scales down to $\Delta t\sim 1$ ms with MCCF and the results, shown 
in the left panel of Fig.~\ref{fig:cyg},
can not be fitted by this model (the solid line in the figure). 
The drifting-blob model (\cite{bot99}) also can explain the observed 
time lags in the Fourier period region above 0.1 s.     
On short time scales, the Comptonization process of the drifting-blob model  
is similar with that of the non-uniform corona model.
Thus the time lag will have similar time scale dependence, i. e. time lag
will decrease with time scale as quickly as the behavior of non-uniform corona
model on short time scales. The drifting blob model
also can not explain the observed time lags of Cyg X-1 on the short time scales.

Poutanen \& Fabian (1999) propose the magnetic flare avalanche model to explain 
the observed time lags of Cyg X-1, and parameterize
the spectral evolution of a magnetic flare and the avalanche process based on
the PCA/$RXTE$ observation on October 23 1996 (ObsID P10241).  
In the right panel of Fig.~6,
we show the measured time lags with the observation data and MCCF technique
(the circles) and the predicted by the model (the solid line).
The predicted time lags can fit the
 measured time lags well in the time scales range between $\sim 4$ 
ms and 1 s, much better than other models, although there seems   
to be overestimated on the shortest timescales. This example shows that
the capability of MCCF for detecting time lags on short time scales
can help us to reveal the underlying physics in high energy process in objects.  
\begin{figure}
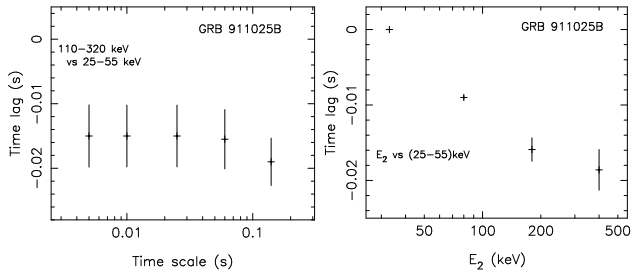

 \includegraphics[width=35mm,angle=270]{f10a.eps}
\includegraphics[width=35mm,angle=270]{f10b.eps}  
 \caption{Hard lags of GRB 911025B measured by MCCF. 
{\sl Left~panel}: Timescale spectra of time lag of (110-320) keV vs. (25-55) 
keV.
{\sl Right~panel}: Energy dependence of time lag of hard photons vs. 
(25-55) keV, averaged for timescales 0.005 s, 0.01 s, 0.25 s, 0.06 s,
and 0.14 s. 
\label{fig:grb}}
\end{figure}

The MCCF is particularly useful  in studying transient processes.  
The BATSE detector has discovered an unexplained phenomenon: a dozen 
intense flashes of hard X-ray and $\gamma$-ray photons of atmospheric 
origin (TGFs) (\cite{fis94}). As all the observed TGFs were 
of short duration (just a few milliseconds), it is difficult to study their 
temporal property by conventional techniques.  With the aid of the preliminary 
MCCF (MCCF$_0$, Eq. (9) in this paper), Feng et al. (2002) 
reveal that for all the flashes with high signal to noise ratio
$\gamma$-ray variations in the low
energy band of 25 - 110 keV relative to the high energy band of $>110$ keV 
are always late in the order of 
$\sim 100$ $\mu$s in the timescale region of $6\times 10^{-6} -
2\times 10^{-4}$~s and pulses are usually wide. 
The above features of energy dependence of
time profiles observed in TGFs support
models that TGFs are produced by upward explosive electrical 
discharges  at high altitude. 

\begin{figure}
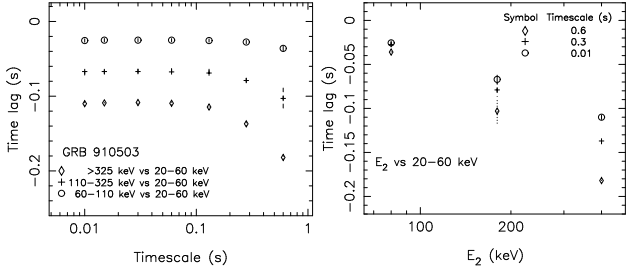

\includegraphics[width=35mm,angle=270]{f11a.eps}
\includegraphics[width=35mm,angle=270]{f11b.eps}  
 \caption{Hard lags of GRB 910503 measured by MCCF. 
{\sl Left~panel}: Timescale spectra of time lag. 
{\it Circle} -- (20-60)keV vs. (60-100)keV; {\it Plus} -- (20-60)keV vs. (110-325)keV;
{\it Diamond} -- (20-60)keV vs. $>325$keV. 
{\sl Right~panel}: Time lag of 20-60 keV photons vs. energy of hard photons.
{\it Circle} -- timescale 0.01 s; {\it Plus} -- timescale 0.3 s;
{\it Diamond} -- timescale 0.6 s. 
\label{fig:grb1}}
\end{figure}

Efforts have been made to measure 
the temporal correlation of two GRB energy bands by the CCF technique  
(e.g. \cite{lin93}; \cite{che95}; \cite{wub00}; \cite{nor02}).
The CCF technique has no necessary sensitivity to make timing analysis for 
weak events. For strong bursts the DISCSC data in BATSE database, 
4-channel light curves with 64 ms time resolution, are usually analyzed, 
but it fails with the traditional ACF and CCF in the case that the existed 
spectral lags comparable or smaller than 64~ms whatever 
how strong the burst is. For short bursts of duration $<$ 2 s, we can
use TTE data to construct high resolution lightcurves, as the four  
lightcurves of time resolution $\delta t=5$ ms for GRB 911025B shown 
in Fig. \ref{fig:grblc}. 
We failed to get statistically meaningful results in spectral lag analysis
for short GRBs by using CCF and MCCF$_0$. The MCCF technique can help
us to reveal spectral lags in strong short bursts. As an example,   
with the lightcurves of GRB 9110258B and MCCF, we calculate the time
lag between two channels at timescale $\Delta t=0.005$ s, 0.01 s, 0.025 s, 
0.06~s, and 0.14~s respectively. In our calculation only partial data having
higher signal to noise ratio recorded during 1.29 s and 1.62 s is used.
The obtained timescale spectrum and energy
dependence of time lags are shown in Fig. \ref{fig:grb}.   
For long $\gamma$-ray bursts, the BATSE Time-to-Spill (TTS) data record the time intervals 
to accumulate 64 counts in each of four energy channels. The TTS data have 
fine time resolution than 64 ms of DISCSC data when the count rate is 
above 1000 cts s$^{-1}$. The TTS data can be binned into equal time bins 
with a resolution of $\delta t \sim 10$ ms    
and our simulations show that from the derived lightcurves the temporal 
and spectral properties with the time resolution $\delta t$ can be reliably 
studied with MCCF for typical GRBs recorded by BATSE.
As an example, the left panel of Fig. \ref{fig:grb1} shows the lag 
spectrum of GRB 910503 detected by BATSE with a duration $\sim 50$ s. 
From the MCCF lag spectra, we can further derive the energy 
dependence of lag at different timescales, shown in the right panel 
of Fig. \ref{fig:grb1}. The results shown in Figs. 10 and 11 indicate that 
MCCF can be used to explore temporal and spectral properties for both
long and short $\gamma$-ray bursts.
    
\section{Discussion}   
The technique for timescale analysis with MCCF is developed from the
standard cross correlation function CCF. For determining CCF 
from unevenly sampled data which are common in astronomical contexts, 
several different methods have been
introduced, i.e., interpolating the data 
between observed points to form a continuous function
(\cite{gas86}), the discrete correlation function (DCF, \cite{ede88}),
evaluating CCF with the discrete Fourier transform (\cite{sca89}),
 and z-transformed discrete correlation function (ZDCF, \cite{ale97}).
Introducing MCCF was motivated by the need of improving the resolution
and sensitivity of the standard correlation analysis. 
Comparing the two definitions, Eq. \ref{mccf} for MCCF and Eq. \ref{ccf} 
for CCF, we can see that MCCF includes more information from the observed 
data than CCF does. That the lag $\tau=k\delta t$ in MCCF has the same 
resolution with the originally observed data but
the resolution of CCF lag is $\Delta t$ and that MCCF is calculated by summing over 
$m=1, 2, \cdots, M_{\Delta t}$ (using all possible lightcurves that can be 
derived from the native data with timescale $\Delta t$) but
CCF only uses one lightcurve for a given time bin $\Delta t$ -- make
MCCF has better resolution and sensitivity in measuring spectral lags.
    
Three different temporal quantities exist in the timescale analysis:
time $t$, time resolution $\delta t$, and time scale $\Delta t$. 
The originally observed data,  based on
which both the frequency analysis and timescale analysis are performed,
is time series $x(t)$ in the time domain with a time resolution $\delta t$.
With frequency analysis, we can derive a frequency spectrum $\Lambda(f)$ 
for a studied characteristic quantity $\Lambda$ 
(in this paper $\Lambda$ is time lag, in other spectral analysis it may
be power density, coherence, or other quantity)
from the observed time series. In frequency analysis, the observed time series 
has to be transformed into the frequency
domain first with the aid of time-frequency transformation, e.g. the Fourier transform,
and the maximum frequency is determined by the time resolution $\delta t$.
The timescale analysis is performed directly in the time domain without any
time-frequency transformation. With timescale analysis we can derive 
a timescale spectrum $\Lambda(\Delta t)$ for the studied quantity $\Lambda$ 
where the argument $\Delta t$, the variation timescale, is a variable similar 
to the frequency $f$ in frequency analysis.
The minimum timescale in timescale analysis is the time resolution 
$\delta t$ of the originally observed data.
To study temporal property on a certain timescale $\Delta t$, 
the used light curves should have a time step equal to the timescale under 
study. With MCCF we can measure any lag greater than the time resolution of observation at any given timescale and
produce a timescale spectrum of time lags.        

There exist two kinds of spectral analysis: frequency
analysis and timescale analysis. Although the Fourier method is a common 
technique to make spectral analysis, it can not replace the timescale analysis
 in the time domain. As any observable physical process always occurs in 
the time domain, a frequency spectrum 
obtained by frequency analysis needs to be interpreted in the time domain.
But a frequency analysis is dependent on a certain time-frequency 
transformation. 
A Fourier spectrum by using Fourier transform with the trigonometric basis 
does not necessarily represent the true distribution of a physical process 
in the time domain. The rms variation vs. 
timescale of a time-varying process may differ substantially 
from its Fourier spectrum, as an example, the Fourier spectrum of a random 
shot series significantly underestimates the power densities 
at shorter timescales (\cite{lit02}).
The present work shows that, like Fourier power spectra, Fourier lag spectra
also always significantly underestimate time lags at short timescales.    
The timescale analysis performed directly in the time domain can derive 
real timescale distribution for quantities characterizing temporal property. 
In comparison with the Fourier technique, timescale spectra of power density 
and time lag from the timescale analysis can more sensitively 
reveal temporal characteristics at short timescales for a complex process.

Welsh (1999) pointed out that the lag determined from the CCF 
should be considered only a characteristic time scale. Care has to be taken 
in interpreting measured lags with a particular physical model.
Most techniques in timing can only treat timescale just 
in a synthetic meaning. For example, a correlation function lag 
of two shot series may caused not only by 
a time displacement between the two series, but also by changes in 
shot shape and intervals between two successive shots. 
Distinguishing different kinds of timescale and physical process 
is obviously helpful to study physics, that should be a goal in 
future development of time domain technique. 
Using a single analysis technique alone is often difficult to definitely 
distinguish the possible processes and compiling different 
results of analysis from different view angles will be helpful.
In comparison with the Fourier technique, the time domain technique has the 
freedom of choosing a proper statistic for a particular purpose. 
Recently Feng, Li \& Zhang (2004) introduced a statistic $w(\Delta t)$ to study 
widths of random shots and diagnosed black hole and neutron star X-ray 
binaries by timing with the new designed statistic.
 
\begin{acknowledgements}
This work is supported
 by the Special Funds for Major State Basic Research Projects and
the National Natural Science Foundation of China. The data analyzed are obtained through the HEASARC on-line service provided 
by the NASA/GSFC. 
\end{acknowledgements}

\end{document}